\documentclass[aps,prl,twocolumn,superscriptaddress,showpacs]{revtex4-2}

\usepackage{amsmath,amssymb}
\usepackage{graphicx}
\usepackage{bm}
\usepackage{xcolor}

\begin{document}

\title{Topological Arrest of Ballooning Modes in Non-Axisymmetric Toroidal Plasmas}

\author{Amitava Bhattacharjee}
\affiliation{Department of Astrophysical Sciences,
Princeton University, Princeton, NJ 08540, USA}
\email{amitava@princeton.edu}

\date{\today}

\begin{abstract}
Why do non-axisymmetric stellarators avoid ballooning
crashes that afflict tokamaks? Three-dimensional geometry
induces Anderson localization of ballooning modes,
converting a global instability into a Ginzburg--Landau
network of isolated wave packets. Global stability reduces
to a percolation problem: below a crtical threshold, instability
is arrested; above it, a crash occurs. This explains
benign stellarator saturation, predicts vulnerability in
quasisymmetric designs, and introduces the critical threshold as a
nonlinear stability metric for reactor optimization,
pending experimental validation.
\end{abstract}

\maketitle

\section{Introduction}
The pursuit of high-pressure plasma confinement in toroidal 
devices is constrained by ballooning instabilities, which can 
drive edge-localized modes (ELMs) in tokamaks~\cite{Strait1994} 
and profile-limiting density collapses in stellarators~\cite{Ohdachi2017}. 
A puzzling asymmetry has emerged: tokamaks are susceptible to 
ELMs that expel significant heat fluxes to the wall, while highly 
non-axisymmetric stellarators such as LHD~\cite{Yamada2010} and 
W7-X~\cite{Grulke2024} operate safely within linearly unstable 
regimes. Understanding this contrast is a central open problem 
in fusion plasma physics.

Ballooning modes are characterized by long-wavelength structures aligned with field lines and short-wavelength structures perpendicular to them. Cuthbert and Dewar~\cite{Cuthbert2000} showed that in general toroidal geometry ballooning eigenfunctions are exponentially localized along field lines and interpreted this behavior in terms of Anderson localization~\cite{Anderson1958}, a well-known phenomenon in condensed matter physics. Redi et al.~\cite{Redi2002} subsequently demonstrated that stellarator ballooning spectra exhibit signatures of quantum chaos. These results suggest that three-dimensional (3D) geometry fundamentally alters the structure of ballooning modes.

In this Letter we propose that Anderson localization converts 
weakly nonlinear ballooning dynamics into a Ginzburg--Landau 
system on a sparse network of localized structures, and that 
global stability is governed by the connectivity of this network. 
We identify a dimensionless percolation parameter $\eta$ and 
argue that global instability can be avoided when $\eta < \eta_c = 
1.128$~\cite{Quintanilla2000}. While the framework requires 
quantitative experimental validation, it suggests that magnetic 
aperiodicity can serve as a topological safety net against global 
ballooning crashes.

\section{Anderson Localization}

In the high-toroidal-mode-number limit the linearized ballooning
equation can be written as~\cite{Cuthbert2000,Dewar1983}
\begin{equation}
\frac{d}{d\theta}\!\left(A(\theta)\frac{d\xi}{d\theta}\right)
  - \bigl[K(\theta) + \lambda N(\theta)\bigr]\xi = 0,
\label{eq:ballooning}
\end{equation}
where $\xi$ is related to the normal component of the plasma
displacement, $A(\theta)$, $K(\theta)$, $N(\theta)$ are functions of
the coordinate $\theta$ along a field line, and $\lambda$ is the linear
eigenfrequency squared. Because $A > 0$, after a standard Liouville transformation, 
this Sturm--Liouville problem can be transformed into Schr\"odinger form
\begin{equation}
\left[\frac{d^{2}}{d\theta^{2}} + E - V(\theta)\right]y = 0,
\label{eq:schrodinger}
\end{equation}
where $y(\theta) = A^{1/2}(\theta)\xi(\theta)$. The effective
potential, $V(\theta)$, depends on equilibrium properties sampled along the field
line and $E$ is the "energy" eigenvalue. Our notation is identical to that of~\cite{Cuthbert2000}.

\paragraph*{Axisymmetric versus 3D geometry.}
In both cases the potential contains a secular magnetic shear term
growing as $|\theta|\to\infty$ that is not periodic. The crucial
distinction lies in the behavior of the geometric coefficients.
For axisymmetry, the curvature, metric coefficients, magnetic field
strength, and pressure-gradient drive are periodic functions of
$\theta$: a field line samples the same geometric sequence repeatedly.
The resulting eigenfunctions are spatially extended Bloch
solutions~\cite{Cuthbert2000,Dewar1983}.

In 3D geometry, by contrast, different field lines on the same flux
surface sample \emph{different} sequences of curvature and metric
coefficients. Along a single field line these coefficients vary
aperiodically as the line winds around the torus. This aperiodic
geometric sampling---not the shear term---is what produces Anderson
localization.

\paragraph*{Transfer-matrix analysis and the Lyapunov exponent.}
Discretizing Eq.~\eqref{eq:schrodinger} on a uniform grid
$\theta_n = n\Delta\theta$ gives the transfer-matrix form
\begin{equation}
\begin{pmatrix}y_{n+1}\\y_n\end{pmatrix}
= T_n\begin{pmatrix}y_n\\y_{n-1}\end{pmatrix},
\label{eq:transfer}
\end{equation}
where 
\begin{equation}
T_n = \begin{pmatrix}2+(\Delta\theta)^2[V(\theta_n)-E]
& -1\\1 & 0\end{pmatrix}.\bigr.
\end{equation}
The Lyapunov exponent of the transfer-matrix product is
\begin{equation}
\gamma = \lim_{N\to\infty}\frac{1}{N}\ln\|T_N T_{N-1}\cdots T_1\|,
\label{eq:lyapunov}
\end{equation}
where $\|\cdot\|$ denotes any convenient matrix norm (the limit is
norm-independent). For 1D Schr\"odinger operators with aperiodic
coefficients, products of transfer matrices generically satisfy
$\gamma > 0$---this is guaranteed under mild assumptions
by the Furstenberg theorem~\cite{Crisanti1993}. A positive Lyapunov
exponent implies exponential localization of eigenfunctions,
\begin{equation}
|y_n| \lesssim e^{-|n-n_0|/\ell},
\label{eq:localization}
\end{equation}
where $n_0$ is the localization center and $\ell \equiv \gamma^{-1}$ is
the localization length. The inequality in Eq.~\eqref{eq:localization}
should be read as a smooth upper-bound envelope; the eigenfunctions
computed numerically by Cuthbert and Dewar exhibit significant fine
structure within the envelope~\cite{Cuthbert2000}.

Furstenberg-type arguments apply under specific assumptions: a given
stellarator field line may be quasiperiodic or nearly integrable, in
which case $\gamma$ should be computed directly from the transfer matrix.
In particular, quasiperiodic potentials---which may arise in
quasisymmetric stellarators where $|B|$ has an approximate continuous
symmetry---can support extended or critical states rather than
Anderson-localized ones.

\section{The Ginzburg--Landau Network}

We derive a time-dependent weakly nonlinear form of the ideal MHD
equations by expanding the fully nonlinear ideal-MHD displacement
equation in Lagrangian coordinates~\cite{Pfirsch1993,Cowley1997,ZhuPoP2007}. The
plasma position is $\mathbf{r}(\mathbf{r}_0,t)=\mathbf{r}_0 +
\bm{\xi}(\mathbf{r}_0,t)$, where $\bm{\xi}$ is the Lagrangian
displacement. Expanding in powers of $\bm{\xi}$ yields
\begin{equation}
\partial_t\xi = \mathcal{L}\xi + \mathcal{N}_2(\xi,\xi)
  + \mathcal{N}_3(\xi,\xi,\xi),
\label{eq:evolution}
\end{equation}
where $\mathcal{L}$ is the linearized ideal-MHD operator and
$\mathcal{N}_2$, $\mathcal{N}_3$ are quadratic and cubic nonlinearities.
The reduction from the second-order ideal-MHD system to the
first-order form~\eqref{eq:evolution} follows from the fact that near
marginal ballooning stability the unstable eigenvalues are purely real
and growing, so the center manifold is one-dimensional and the
dynamics project onto a non-oscillatory equation for
each mode amplitude~\cite{Crawford1991}. Near marginal stability the
perturbation on this reduced manifold is
\begin{equation}
\xi(\theta,t) = \sum_j a_j(t)\,\phi_j(\theta),
\label{eq:expansion}
\end{equation}
where $\phi_j(\theta)$ are the weakly unstable and Anderson-localized
eigenfunctions. As derived in the Appendix and the End Matter, the
amplitude equations take the Ginzburg--Landau form
\begin{equation}
\dot{a}_j = \gamma_j a_j - \mu_j a_j^3
  + \sum_{k\neq j} J_{jk}\,a_k,
\label{eq:GL}
\end{equation}
where the linear growth rate $\gamma_j$ (not to be confused with the
Lyapunov exponent) and the cubic coefficient $\mu_j > 0$
are mode-specific. The inter-mode coupling decays exponentially
with separation:
\begin{equation}
J_{jk} \sim J_0\,e^{-d_{jk}/\ell}.
\label{eq:coupling}
\end{equation}
Here each localized packet is labeled by a field-line label $\alpha_j$
and a ballooning-space localization center $\theta_j$; its projection
onto the flux surface defines a point $x_j$, and $d_{jk}$ is the metric distance on the surface. The absence of
the quadratic term $\mathcal{N}_2$ from Eq.~\eqref{eq:GL} is a central result whose
derivation is given in the End Matter.

The Ginzburg--Landau network~\eqref{eq:GL} naturally contains the
intermediate nonlinear regime first identified by Zhu et al.~\cite{ZhuPRL2006} and subsequently developed for
toroidal ballooning modes~\cite{ZhuPoP2007,ZhuPRL2009}. Prior to
cubic saturation, when $\mu_j a_j^2 \ll \gamma_j$, each mode
amplitude continues to grow exponentially at the linear rate $\gamma_j$
with its spatial structure unchanged. In 3D geometry the intermediate
regime is followed by cubic saturation. In axisymmetric geometry,
sufficiently close to marginal stability, the quadratic
drive~\cite{Cowley1997} can in principle follow the intermediate
regime and produce explosive growth; whether it does so in practice
depends on proximity to marginality and the absence of competing
saturation mechanisms.

Two localized structures interact effectively only when the coupling
$J_{jk}$ exceeds a characteristic dynamical scale $\sim\gamma_j$,
the linear growth rate of the mode. This defines an effective
interaction radius
\begin{equation}
R_{*} = \ell\ln\!\left(\frac{J_0}{\gamma_j}\right).
\label{eq:Rstar}
\end{equation}
Here $R_{*}$ is determined entirely by equilibrium and linear stability properties
with no free parameters. On a flux surface the centers $\theta_j$ of
unstable localized packets are distributed with areal density
$\rho(\beta)$, which increases as the pressure gradient exceeds the
marginal value. The set of packets and their interaction radii defines
a random geometric graph on the flux surface. A standard result of
continuum percolation theory~\cite{Quintanilla2000} is that such a
graph develops a surface-spanning connected cluster when the
dimensionless parameter
\begin{equation}
\eta(\beta) = \rho(\beta)\,\pi R_{*}^{2}(\beta)
\label{eq:eta}
\end{equation}
exceeds the critical value $\eta_c = 1.128$. Recent M3D-C$^1$
simulations of W7-X plasmas show regimes of benign saturation
($\beta < 5\%$) and crash-like behavior in configurations with near-zero
magnetic shear~\cite{Zhou2024,Wright2024}. In the present framework
these behaviors correspond naturally to subcritical and supercritical
connectivity of localized ballooning structures.

We note that this percolation threshold provides a necessary but not
sufficient condition for global instability. The present framework
treats local growth rates as approximately uniform and coupling phases
as random. In practice, heterogeneity in growth rates could produce mode clustering that locally
restores quadratic coupling even in a globally subcritical network.
Conversely, phase interference among coupling terms could raise the
effective threshold above $\eta_c$. A fully quantitative stability
criterion would require accounting for these effects, which we leave
for future work. The percolation threshold on the toroidal flux surface
(topologically a 2-torus) differs from the flat-plane value by
finite-size corrections that are small for large $N$. Furthermore, if the
interaction region is anisotropic (elongated along field lines), the
relevant percolation threshold for elliptical objects differs from
1.128, though the qualitative phase-transition structure is preserved.

To connect $\eta$ with experiment and simulation, Fig.~1 shows the
topological connectivity for three characteristic magnetic geometries,
and Fig.~2 provides both a schematic regime map and the theoretical scaling.

\begin{figure}[t]
\includegraphics[width=\columnwidth]{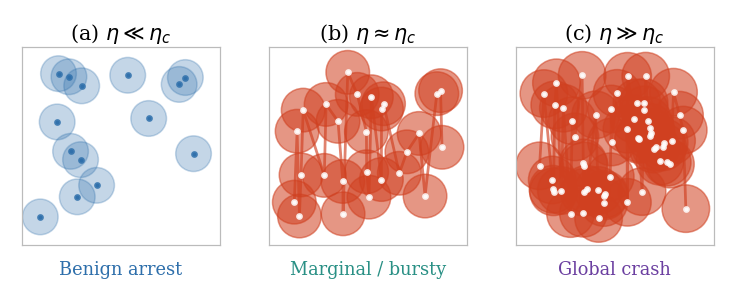}
\caption{Schematic illustration of the three topological
connectivity regimes on a toroidal flux surface. Each disk represents
a localized ballooning mode packet; disk radii are chosen to give
$\eta$ values representative of each regime and are not derived from
equilibrium calculations. \textbf{(a)}~Subcritical ($\eta\ll\eta_c$):
high magnetic aperiodicity produces short localization lengths and
sparse, non-percolating clusters; global synchronization is prevented
and instability saturates benignly. \textbf{(b)}~Near-critical
($\eta\approx\eta_c=1.128$): a surface-spanning cluster path
(highlighted) intermittently forms, enabling bursty, intermittent
MHD activity. \textbf{(c)}~Supercritical ($\eta\gg\eta_c$): the
spanning cluster persists, phase-locking modes globally and enabling
the explosive growth characteristic of a major profile crash.}
\label{fig:connectivity}
\end{figure}

For the schematic regime map (Fig.~2, left panel), device positions
are based on qualitative reasoning from published results rather than
direct extraction of $\rho$ and $R_*$. W7-X simulations show benign
nonlinear saturation up to $\beta\approx 5\%$~\cite{Zhou2024},
consistent with $\eta\ll\eta_c$; LHD exhibits bursty MHD activity
and core density collapse events near its ballooning
threshold~\cite{Ohdachi2017}, consistent with $\eta\lesssim\eta_c$;
and DIII-D inter-ELM measurements of radial correlation lengths of
density fluctuations ($0.5$--$1$~cm) from reflectometry and BES
diagnostics~\cite{Yan2011,Rhodes1995,Barada2021}, combined with
filamentary scales in W7-X and LHD~\cite{Buzas2020,Tanaka2012},
provide proxy estimates placing DIII-D in a moderately supercritical
regime. We repeat for emphasis that no published study directly reports $\rho$ or
$R_*$ in the percolation sense; the values in Fig.~2 are
order-of-magnitude estimates with uncertainty bands spanning roughly
one order of magnitude. The right panel of Fig.~2 shows the
theoretical scaling $\eta\sim(a/\ell)^2[\ln(J_0/\gamma)]^2$, derived in the End Matter,
as a function of the ratio $a/\ell$ of machine size to localization length,
with device positions placed at qualitative estimates of this ratio.
This scaling makes explicit that axisymmetric geometry ($\ell\to\infty$,
$a/\ell\to 0$) formally gives $\eta\to\infty$ for any finite $\beta$
above marginal stability.

A quantitative determination of $\eta$ would require: (i) computing
$\gamma$ (and hence $\ell$) from the transfer-matrix
Eq.~\eqref{eq:transfer} evaluated along field lines of each
equilibrium reconstruction; (ii) measuring $\rho = N/A$ from the
linear stability boundary as a function of field-line label $\alpha$,
i.e.\ the fraction of field lines on the flux surface that are
ballooning-unstable at a given $\beta$; and (iii) determining $R_*$
via Eq.~\eqref{eq:Rstar} with $\gamma_j$ from linear stability
calculations. Steps (i) and (ii) are computable from
equilibrium reconstructions and would convert Fig.~2 from a schematic
regime map into a parameter-free theoretical prediction testable
against simulation. Step (iii) could be validated experimentally
from spatiotemporal correlation functions of coherent
fluctuation structures in imaging or probe data.

\begin{figure*}[t]
\includegraphics[width=\textwidth]{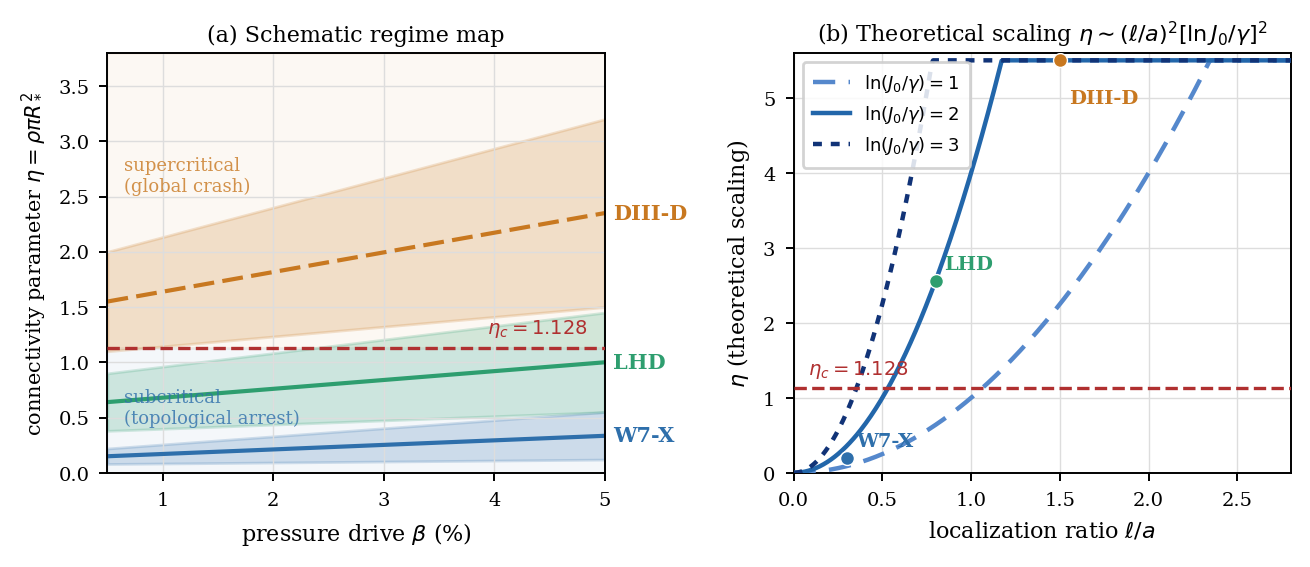}
\caption{\textbf{(a)}~Schematic estimates of the connectivity parameter
$\eta=\rho\pi R_*^2$ for representative operating regimes. Bands reflect
order-of-magnitude uncertainty; no published study directly reports $\rho$
or $R_*$ in the percolation sense. Device positions are based on the
qualitative reasoning described in the text: W7-X (M3D-C$^1$ simulations
show benign saturation to $\beta\approx5\%$~\cite{Zhou2024}); LHD (bursty
MHD activity and density crashes near the ballooning
threshold~\cite{Ohdachi2017}); DIII-D inter-ELM (radial correlation
lengths $0.5$--$1$~cm from reflectometry and BES~\cite{Yan2011,Rhodes1995,
Barada2021}, filamentary scales from~\cite{Buzas2020,Tanaka2012}). Dashed
line: continuum percolation threshold $\eta_c=1.128$~\cite{Quintanilla2000}.
\textbf{(b)}~Theoretical scaling $\eta\sim(\ell/a)^2[\ln(J_0/\gamma)]^2$
for three values of the dimensionless coupling ratio. Device positions are
qualitative estimates of $\ell/a$; axisymmetric geometry ($\ell\to\infty$,
$a/\ell\to0$) gives $\eta\to\infty$.}
\label{fig:phasediagram}
\end{figure*}

\section{Conclusions}

We have demonstrated that 3D geometry converts ballooning dynamics from
a single global mode into a Ginzburg--Landau system on a network of
Anderson-localized states. Nonlinear stability is therefore a
connectivity property of this network. Localization is a
\emph{consequence} of the Furstenberg theorem applied to the
aperiodic ballooning potential in 3D geometry. The
Anderson mechanism is absent in axisymmetric geometry, where extended
Bloch eigenfunctions support the explosive quadratic nonlinearity of
the detonation theory~\cite{Cowley1997}.

In the presence of symmetry-breaking imposed by resonant magnetic
perturbations in a tokamak, the framework predicts that sufficient
magnetic aperiodicity could push $\eta$ below $\eta_c$, providing a
topological interpretation of ELM suppression. This is a prediction of
the theory amenable to experimental test.

Recent work suggests that in quasisymmetric equilibria the magnetic
field strength along a flux surface may possess an underlying structure
associated with integrable systems such as the KdV
equation~\cite{Sengupta2025}. The spectrum of the associated
Schr\"odinger-like operator for KdV-type (reflectionless) potentials
consists of extended, transmitting states rather than Anderson-localized
ones---precisely the case where the Furstenberg
theorem does not apply. This connection strengthens the conjecture that stellarators
with near-perfect quasisymmetry may prove as vulnerable to nonlinear
ballooning modes as axisymmetric tokamaks. Recent M3D-C$^1$ simulations
of Landreman--Buller--Drevlak-type quasisymmetric equilibria showing
sustained ballooning instability ~\cite{Wright2024} seem qualitatively
consistent with this prediction , and motivate dedicated linear
calculations of the Lyapunov exponent for these configurations.

Density crashes have been observed in LHD by Ohdachi et al.~\cite{Ohdachi2017},
showing a ballooning mode in the gradient region driving a
beta-limiting collapse. This provides possible evidence for the Anderson localization mechanism: it is spatially localized, destabilized by 3D geometry, and appears in a region of globally negative magnetic shear where axisymmetric ballooning theory predicts stability. Within the present framework, this mode is a localized packet that survives stabilizing global shear because the aperiodic 3D potential creates an isolated potential well along a specific field line — a phenomenon that cannot occur in axisymmetric geometry. The intermittent, bursty character of the density collapse is consistent with the near-critical regime $\eta \approx \eta_c$, in which the spanning cluster forms and dissolves repeatedly as the pressure profile oscillates about the percolation threshold.

The percolation parameter $\eta$ could serve as an additional
optimization target in large-scale stellarator design codes alongside
existing neoclassical and linear-MHD metrics---providing a nonlinear
stability criterion that is in principle computable from equilibrium
and linear stability data. By strategically retaining a degree of
geometric aperiodicity, it may be possible to design reactors that are
topologically immune to global ballooning crashes, trading near-perfect
quasisymmetry for robust nonlinear stability.

\begin{acknowledgments}
We thank N.~Bohlsen, A.~Brown, S.~Buller, I.~Dodin, P.~Helander, N.~Nikulsin, H. Qin, and W.~Sengupta for helpful
discussions. We acknowledge the use of large language models (Anthropic (Claude), ChatGPT, and Gemini)
for assistance with drafting, organizing ideas, and navigating the literature. 
The author is responsible for all scientific content and
conclusions. This research was supported by a grant from the Simons Foundation/SFARI (560651, AB) and
DOE Award No.\ DE-SC0024548 (until March 31, 2025). We are grateful to
Princeton University for supporting our research during the suspension
period of our DOE Award.
\end{acknowledgments}

\appendix
\section{Appendix: Derivation of the Ginzburg--Landau Equation}

We describe how the weakly nonlinear Ginzburg--Landau evolution follows
from the fully nonlinear Lagrangian ideal-MHD displacement equation.
In Lagrangian variables, ideal MHD admits an exact equation of
motion~\cite{Pfirsch1993,Cowley1997,ZhuPoP2007}
\begin{equation}
\rho_0(\mathbf{a})\partial^2_t\bm{\xi} = \mathbf{F}[\bm{\xi}],
\label{eq:exactMHD}
\end{equation}
where $\mathbf{F}[\bm{\xi}]$ is a nonlinear force functional. Expanding
about equilibrium gives
\begin{equation}
\mathbf{F}[\bm{\xi}] = \mathcal{L}\bm{\xi}
  + \mathcal{Q}(\bm{\xi},\bm{\xi})
  + \mathcal{C}(\bm{\xi},\bm{\xi},\bm{\xi}) + \cdots,
\label{eq:expansion_F}
\end{equation}
where $\mathcal{L}$ is the linear ideal-MHD force operator and
$\mathcal{Q}$, $\mathcal{C}$ are quadratic and cubic nonlinearities.
Near marginal ballooning stability the spectrum of $\mathcal{L}$
contains a finite set of weakly unstable purely real eigenvalues. The
corresponding unstable center manifold is 1D, and
standard center-manifold analysis~\cite{Crawford1991} reduces
Eq.~\eqref{eq:exactMHD} to the first-order
system~\eqref{eq:evolution}. Projecting onto the adjoint eigenfunctions
$\phi^\dagger_j$ yields the amplitude equations
\begin{equation}
\dot{a}_j = \gamma_j a_j + \sum_{kl}c^{(2)}_{j;kl}a_ka_l
  + \sum_{klm}c^{(3)}_{j;klm}a_ka_la_m + \cdots,
\label{eq:amplitude_general}
\end{equation}
where $c^{(2)}_{j;kl} = \langle\phi^\dagger_j,
\mathcal{Q}(\phi_k,\phi_l)\rangle$ and analogously for $c^{(3)}$. The
suppression of the quadratic coupling in 3D geometry is demonstrated in
the End Matter. Retaining the leading contributions gives
Eq.~\eqref{eq:GL} with $J_{jk}\sim e^{-d_{jk}/\ell}$.

\section*{References}

\section*{End Matter}

\subsection*{Suppression of quadratic couplings}

Localization does not make $c^{(2)}_{j;kl}$ vanish identically, but
it does make most couplings exponentially small. If packet envelopes
satisfy $|\phi_j|\lesssim e^{-d(\theta,\theta_j)/\ell}$, then
\begin{equation}
|c^{(2)}_{j;kl}|\lesssim C_2\,e^{-D_{jkl}/\ell},
\label{eq:overlap}
\end{equation}
where $D_{jkl}$ measures the mutual separation of the three packets.
Off-diagonal quadratic couplings are therefore exponentially
suppressed in the strongly localized regime.

The self-coupling $k=l=j$ is not suppressed by overlap. In
stellarator-symmetric equilibria, where $V(\theta)$ is even about
the localization center $\theta_j$, the ground-state eigenfunction
is even: $\phi_j=f_j(\theta-\theta_j)$, $f_j(-u)=f_j(u)$,
$f_j\to0$ exponentially. Since $Af_j^2f_j'=\tfrac{A}{3}(f_j^3)'$,
integration by parts gives $c^{(2)}_{j;jj}=-\tfrac{1}{3}\int\! A'f_j^3\,d\theta
\sim O(\ell/L_A)$. Here $L_A$ is
the variation scale of $A$ and $b_j\equiv c^{(2)}_{j;jj}\sim O(\ell/L_A)$ is the self-quadratic
coefficient. Neglecting it relative to the cubic requires
$|b_j|\ll\mu_j\sqrt{\gamma_j/\mu_j}$ at the saturation amplitude
$a_j\sim\sqrt{\gamma_j/\mu_j}$.

The key distinction from axisymmetric geometry is \emph{phase coherence}. In axisymmetric geometry the
Bloch eigenfunction $\phi\sim P(\theta)e^{-\hat{s}^2\theta^2/2\ell_s^2}$
carries a periodic phase structure that supports a two-step harmonic
cascade: the primary mode at $m$ drives a second harmonic at $2m$, 
which feeds back to drive the detonation mechanism~\cite{Cowley1997}. Anderson
localization destroys this extended phase coherence, converting the
quadratic interaction into short-ranged incoherent couplings.

For quasi-randomly distributed centers, the $N$ terms in $\mathcal{Q}_j$ contribute with quasi-random
signs. By the Brownian-walk argument, $|\mathcal{Q}_j|\sim\sqrt{N}$
while the collective cubic scales as $N$:
\begin{equation}
\frac{|\mathcal{Q}_j|}{\sum_j\mu_ja_j^3}
\sim\frac{1}{\sqrt{N}}
\;\xrightarrow{N\to\infty}\;0.
\label{eq:ratio}
\end{equation}
The cubic Ginzberg-Landau network~\eqref{eq:GL} is therefore the correct leading dynamics
in the large-$N$, strongly-localized regime.

\subsection*{Percolation parameter scaling}

With $\rho=f(\beta)/a^2$ (number of packets $\sim f$ over machine
area $a^2$) and $R_*=\ell\ln(J_0/\gamma)$:
\begin{equation}
\eta=\rho\pi R_*^2
\sim \pi f\!\left(\frac{\ell}{a}\right)^{\!2}
\ln^2\!\!\left(\frac{J_0}{\gamma}\right).
\label{eq:eta_scale}
\end{equation}
As $\ell\to\infty$ (axisymmetric limit), $\eta\to\infty$ for any
$\beta>\beta_c$, consistent with the explosive character of tokamak
ballooning instabilities and with Fig.~2(b). Near marginal $\beta_c$,
$f\sim\epsilon\equiv(\beta-\beta_c)/\beta_c$, giving critical overshoot
$\epsilon_c\sim a^2\eta_c/[\pi\ell^2\ln^2(J_0/\gamma)]$; for
$\ell/a\sim0.1$--$0.3$ and $\ln(J_0/\gamma)\sim2$--$3$,
$\epsilon_c\sim0.03$--$0.10$, consistent with observed
ballooning-limit windows.

\end{document}